\documentclass[12pt]{iopart}
\usepackage{iopams}

\usepackage[english]{babel}

\usepackage{amsfonts}
\usepackage{amssymb}
\usepackage{amsthm}
\usepackage{amscd}
\usepackage{epsfig}

\usepackage[latin1]{inputenc}
\usepackage{a4wide}

\def\p{\partial}

\def\d{\delta}

\def\k{\kappa}

\newcommand{\Sc}{Schr\"o\-din\-ger }
\newcommand{\eqref}[1]{(\ref{#1})}

\newtheorem{Th}{Theorem}
\newtheorem{lemma}{Lemma}

\newcommand{\diag}{\mbox{\rm diag}}

\begin{document}

\title[Eigenphase preserving two-channel SUSY transformations]%
{Eigenphase preserving two-channel SUSY transformations}

\author{Andrey M Pupasov$^{1, 2}$\footnote{Boursier de l'ULB.},
Boris F Samsonov$^1$, Jean-Marc Sparenberg$^2$ and Daniel Baye$^2$}

\address{$^1$ Physics Department, Tomsk State University, 36 Lenin Avenue,
634050 Tomsk, Russia}

\address{$^2$ Physique Quantique, C.P.\ 229, Universit\'e Libre de Bruxelles,
B 1050 Bruxelles, Belgium}

\eads{\mailto{pupasov@phys.tsu.ru}, \mailto{samsonov@phys.tsu.ru},
\mailto{jmspar@ulb.ac.be}, \mailto{dbaye@ulb.ac.be}}

\begin{abstract}
We propose a new kind of supersymmetric (SUSY) transformation in the case of
the two-channel scattering problem with equal thresholds,
for partial waves of the same parity.
This two-fold transformation is based on two imaginary factorization energies with opposite signs and with mutually conjugated factorization solutions.
We call it an eigenphase preserving SUSY transformation
as it relates two Hamiltonians,
the scattering matrices of which have identical eigenphase shifts.
In contrast to known phase-equivalent transformations,
the mixing parameter is modified by the eigenphase preserving transformation.
\end{abstract}


\centerline{\today}

\section{Introduction}
The present work is a continuation of our previous investigations on
supersymmetric (SUSY) transformations
applied to coupled-channel problems with equal thresholds \cite{pupasov:09}.
Our main aim here is to present a method based on SUSY transformations, which allows
to construct potentials with given scattering properties, i.e.,
to solve an inverse scattering coupled-channel problem.

There are several approaches to this problem based on the
Gelfand-Levitan-Marchenko methods \cite{gelfand:51,marchenko:55}.
In particular, Newton, Jost and Fulton
\cite{jost:55}-\cite{newton:57}
generalized the Gelfand-Levitan method
and solved the corresponding integral equations in the case
of two channels and rational scattering ($S$) matrices.
Exactly solvable coupled-channel potentials
obtained by this technique
may be used for describing the neutron-proton scattering.
In particular, in this way, Newton and Fulton \cite{newton:57}
 constructed a three-parameter phenomenological
neutron-proton potential fitting low-energy scattering data.
It would be interesting to extend this result by enlarging the number of parameters to fit scattering data on a wider energy range;
however, the method based on integral transformations is rather involved
and therefore quite difficult to generalize.
Using the Marchenko equation, the results of Newton and Fulton were
nevertheless reproduced and improved by von Geramb \etal \cite{geramb}.
A review of the present
state of the art in the inverse scattering methods
may be found in \cite{ISM1}-\cite{ISM2}.

Our hope that the SUSY technique may be efficient for the
multichannel \Sc equation
is based on the well known equivalence
between SUSY transformations and the integral transformations of the
inverse scattering method
for single-channel problems
\cite{sukumar:85}-\cite{baye:04}.
Due to this equivalence,
one can use chains of first-order SUSY operators
(also referred to as first-order Darboux differential operators \cite{samsonov:95})
for constructing a Hamiltonian with given scattering properties
\cite{sparenberg:97a,samsonov:03}.
This approach to the scattering inversion
is more efficient \cite{baye:04} just because of the
differential character of the transformation.
There are several papers
devoted to supersymmetric transformations for multichannel
problems \cite{amado:88a}-\cite{sparenberg:06}
(see also
\cite{samsonov:07}-\cite{sparenberg:08}
for additional motivations and physical applications).
Arbitrary chains of first-order SUSY transformations in the case of a matrix \Sc equation
are studied in \cite{pechericyn:04m}.
There, a matrix generalization of the well-known Crum-Krein formula is obtained.
Another important ingredient of the supersymmetric inversion technique
are the phase-equivalent SUSY transformations,
which are based on two-fold, or second-order, differential operators.
These are described in \cite{baye:87}-\cite{samsonov:02}
for the single-channel case
and in \cite{sparenberg:97,Leeb} for the coupled-channel case.
Such transformations keep the scattering matrix unchanged
and simultaneously allow to reproduce given bound state properties.

It should be noted that methods
based on a direct generalization of the SUSY technique
to the multichannel case
are not able to provide an easy control of the scattering properties
 for all channels simultaneously.
For instance, in the two-channel case, the $S$-matrix is parametrized
by the eigenphase shifts $\delta_1(k)$, $\delta_2(k)$ and mixing
parameter $\epsilon(k)$,
where $k$ is the wave number.
Usual SUSY transformations modify these three quantities in a complicated way,
which makes their individual control difficult.
 We believe that this is the reason why
SUSY transformations did not find
a wide application to multichannel scattering inversion.

In the present paper, we propose a two-fold SUSY transformation
which allows us to modify $\epsilon(k)$ only,
while keeping $\delta_1(k)$ and $\delta_2(k)$ unchanged.
We call such a transformation {\it eigenphase preserving}.
It is necessary to stress the difference between this new kind of transformation
and the well-known phase-equivalent transformations mentioned above.
A phase-equivalent transformation does not modify the scattering matrix at all,
whereas the eigenphase preserving transformation modifies the mixing between channels.
An important consequence of that is the possibility
to use single channel SUSY transformations to
fit experimental values of the eigenphase shifts.
Afterwards, the mixing parameter can be fitted without further modification of
the eigenphase shifts.
Thus, the main advantage of our approach
consists in splitting the inversion problem into two independent parts:
(i)~fitting eigenphase shifts to experimental values
independently for each channel and
(ii)~fitting the mixing parameter between these channels.
To solve the first problem, one can use the single-channel tools mentioned above.
In the present work, we propose an elegant solution to the second problem.

In what follows, we will use definitions and notations introduced
in our previous paper \cite{pupasov:09}, where a first order
coupling SUSY transformation is analyzed in details.
Nevertheless, in section 2, we recall some basic formulae
necessary in the next sections.
In section 3, we describe the new two-fold SUSY transformation
and prove our main result
that this transformation preserves the eigenphase shifts.
A simple illustrative example of an exactly-solvable coupled-channel potential
with a given scattering matrix is presented in section 4.
In the conclusion, we discuss possible applications of the presented
method and formulate some possible lines of future investigations.

\section{Two channel scattering with equal thresholds}

Consider the two component radial \Sc equation \cite{taylor:72,newton:82}
\begin{equation}\label{se1}
H_0\psi_0(k,r)=k^2\psi_0(k,r),\quad r\in[0,\infty)
\end{equation}
with the Hamiltonian
\begin{equation}
H_0=-\mathbf{1}\frac{d^2}{d r^2}+ V^\mathrm{int}_0(r)+
l(l+\mathbf{1})r^{-2},\qquad l=\diag(l_1,l_2)\,.
\end{equation}
Here $\mathbf{1}$ is the $2\times 2$ identity matrix
and the interaction potential $V^\mathrm{int}_0(r)$ is a real and symmetric matrix,
exponentially decreasing at large distances.
We will consider the case of two partial waves $l_1$ and $l_2$
with identical parity,
\begin{equation}\label{l2l1}
 l_2 = l_1 + 2 m, \qquad m \in \mathbb{Z}.
\end{equation}
For the sake of convenience,
we combine interaction $V^\mathrm{int}_0(r)$ and the centrifugal
term into a single potential  matrix
$V_0(r)=V^\mathrm{int}_0(r)+l(l+\mathbf{1})r^{-2}$.
To characterize this potential near the origin,
we use a matrix singularity index $\nu$.
Matrix $\nu$ is determined by the asymptotic
behaviour of the potential near the origin,
\begin{equation}\label{Vzero}
V_0(r\rightarrow 0)=\nu(\nu+\mathbf{1})r^{-2}+{\rm O}(1)\,.
\end{equation}
Below, only potentials with singularity index
being a diagonal matrix with integer entries $\nu=\diag(\nu_1,\nu_2)$ and
$\nu_j \ge l_j$ { are considered}.
We will call such potentials physical
and restrict ourselves to SUSY transformations that produce
physical potentials.

As usual, the Jost solutions $f_0(k,r)$ are defined as matrix
solutions of \eqref{se1} with exponential asymptotic behaviour at large distances \cite {taylor:72,newton:82}.
In what follows, we will need a more detailed asymptotic behaviour of these solutions;
it is given by the asymptotic behaviour of the Bessel functions of the third kind, $H_{l+\frac{1}{2}}^{(1)}(z)$, also called the first Hankel functions
(see \cite{Beitman} for a definition).
At large distances, the Jost solution thus behaves like the corresponding solution for the free particle
\begin{equation}\label{jsas}
f_0(k,r\rightarrow\infty)\to {\rm
diag}\left[h_{l_1}(kr),h_{l_2}(kr)\right]\,,\qquad
h_l(z)=i^{l+1}\left(\frac{\pi z}{2}\right)^{\frac{1}{2}}
H_{l+\frac{1}{2}}^{(1)}(z)
\end{equation}
with
\begin{equation}\label{asyh}
h_l(z\rightarrow\infty)=e^{iz}\left(1+\frac{i\Lambda}{2z}+{\rm
o}(z^{-1})\right),\qquad \Lambda=l(l+1).
\end{equation}

A special linear combination of the Jost solutions gives
the regular solution
 \begin{equation}\label{rajs}
\varphi_0(k,r)=\frac{i}{2k}\left[f_0(-k,r)F_0(k)-f_0(k,r)F_0(-k)\right],
\end{equation}
 \begin{equation}\label{rajs-bb}
\varphi_0(k,r\rightarrow 0)\to
{\diag}\left(\frac{r^{\nu_1+1}}{(2\nu_1+1)!!},\frac{r^{\nu_2+1}}{(2\nu_2+1)!!}\right)\,,
\end{equation}
where matrix $F_0(k)$ is the so-called Jost matrix.

To construct eigenphase preserving transformations,
 we need
 solutions of the \Sc equation  (\ref{se1}) with a special behaviour both
 at large distances and near the origin.
 Thus, we first prove that the necessary solutions exist.

\begin{lemma}\label{l-1}
For any momentum $k$ such that {\rm Im} $k>0$, {\rm det} $F_0(k)\neq 0$,
and for any constants $c_{1,2},d_{1,2}\in \mathbb{C}$, there exist two vector solutions
$\vec u(k,r)$
and $\vec v(k,r)$
of the \Sc equation \eqref{se1}
which behave  at large distances as
\begin{equation}\label{rajs-vas}
\vec u(k,r\rightarrow\infty)={\rm e}^{-ikr}(c_1,c_2)^T(1+{\rm o}(1))\,,
\end{equation}
\begin{equation}\label{js-vas}
\vec v(k,r\rightarrow\infty)={\rm e}^{ikr}(d_1,d_2)^T(1+{\rm o}(1))\,,
\end{equation}
and near the origin as
\begin{equation}\label{rajs-vbb}
\vec u(k,r\rightarrow 0)=(a_1r^{\nu_1+1},a_2r^{\nu_2+1})^T(1+{\rm o}(r))\,,
\end{equation}
\begin{equation}\label{js-vbb}
\vec v(k,r\rightarrow 0)=(b_1r^{-\nu_1},b_2r^{-\nu_2})^T(1+{\rm o}(r))\, ,
\end{equation}
where
$a_{1,2},b_{1,2}\in \mathbb{C}$.
\end{lemma}

\begin{proof}
To obtain the behaviour (\ref{js-vas}), $\vec v(k,r)$ can be expressed in terms of the Jost solution
\begin{equation}\label{js-vas-1}
\vec v(k,r)=f_0(k,r)(d_1,d_2)^T.
\end{equation}
Formula \eqref{js-vbb} follows
from the behaviour of the Jost solution near the origin
(see, e.g., \cite{taylor:72}).

Taking into account that Im $k>0$, one gets from \eqref{rajs}
\begin{equation}\label{rajs-as}
\varphi_0(k,r\rightarrow\infty) \rightarrow \frac{i}{2k}f_0(-k,r)F_0(k)\,.
\end{equation}
Here, we omit the second term in \eqref{rajs} since it
becomes negligible at large distances with respect to the
first term.
Thus, solution $\vec u(k,r)$ may be obtained as
\begin{equation}\label{rajs-vas-1}
\vec u(k,r)=\frac{2k}{i}\varphi_0(k,r)F_0^{-1}(k)(c_1,c_2)^T.
\end{equation}
Formula \eqref{rajs-vbb} follows from \eqref{rajs-bb}.
\end{proof}

The $2 \times 2$ scattering matrix $S_0(k)$
is expressed in terms of the Jost matrix as
\begin{equation}\label{SdefI}
S_0(k)=e^{il\frac{\pi}{2}}F_0(-k)F_0^{-1}(k)e^{il\frac{\pi}{2}}\,.
\end{equation}
Being unitary and symmetric, $S_0(k)$ can be diagonalized
by an energy dependent orthogonal matrix $R_0(k)$
\begin{equation}\label{S0diag}
R_0^T(k) S_0(k) R_0(k)= \diag \left(e^{2i\delta_{0;1}(k)},e^{2i\delta_{0;2}(k)}\right),
\end{equation}
where $\delta_{0;j}$ are the eigenphase shifts and
the angle $\epsilon_0$ entering matrix $R_0$
is called the mixing angle
\begin{equation}\label{par-def}
\qquad R_0(k)= \left(
\begin{array}{cc}
 \cos\epsilon_0(k) & -\sin\epsilon_0(k)
\\ \sin\epsilon_0(k) &\cos\epsilon_0(k)
\end{array}
\right).
\end{equation}
Note that an opposite sign definition for the mixing angle could
have been chosen; moreover, the order of the eigenphase shifts is
arbitrary: exchanging them while adding $\pm\pi/2$ to the mixing
angle keeps the scattering matrix unchanged. In the next section,
the eigenphase preserving SUSY transformations
 are defined.

\section{Eigenphase preserving two-fold SUSY transformations}

\subsection{Two-fold SUSY transformations}

Two-fold SUSY transformations lead to a number of interesting
quantum models with unusual properties \cite{andrianov:95sdt}.
In particular, the corresponding superalgebra
is nonlinear.
It is natural to consider the two-fold SUSY transformation
of the \Sc equation \eqref{se1}
 as a chain of usual (i.e.\ one-fold) SUSY transformations.
One-fold transformations
for coupled-channel
\Sc equations  were introduced in \cite{amado:88a}.
Their generalization,
which allows to introduce a coupling between channels,
was given in \cite{pupasov:09}.
The additional requirement that the transformed potential be
physical was shown to result in a strong constraint on the transformation parameters.
 The case
of two transformations is less restrictive since the intermediate
Hamiltonian may be chosen unphysical. In particular, one may use
as transformation functions complex-valued solutions of the \Sc
equation corresponding to complex factorization constants.
As we show below, a chain of two such transformations may preserve the
eigenphase shifts.

The chain of two SUSY transformations,
$H_0\rightarrow H_1\rightarrow H_2$,
emerges from the following intertwining relations:
\begin{equation}\label{ir1}
L_{1}H_0=H_1L_{1}\,,\qquad L_{2}H_1=H_2L_{2}\,,
\end{equation}
where the operators $L_j$ map solutions of the \Sc equations to each other as $\psi_1=L_1 \psi_0$ and $\psi_2=L_2 \psi_1$.
These operators can be combined into an operator $L$ defining the two-fold SUSY transformation
\begin{equation}\label{ir2}
LH_0=H_2L\,,\qquad L=L_2 L_1\,,
\end{equation}
directly mapping solutions of the initial \Sc
equation to solutions of the trans\-formed \Sc equation as
$\psi_2=L\psi_0$.

The operators $L_j$ are first-order differential operators,
\begin{eqnarray}\label{ct}
L_{1} = w_1(r)-\partial_r\,, \qquad
L_{2} = \tilde{w}_2(r)-\partial_r\,.
\end{eqnarray}
We use the standard notation for the superpotentials
\begin{eqnarray}\label{spts12}
w_j(r) & = & u_j'(r)u_j^{-1}(r)\,,\qquad j=1,2\,, \\
\label{w2tilde}
\tilde{w}_2(r) & = & \tilde{u}'_2(r)\tilde{u}_2^{-1}(r)\,,
\end{eqnarray}
which are expressed in terms of the matrix factorization solutions $u_j$ and $\tilde u_2=L_1 u_2$.
These solutions satisfy the following \Sc equations:
\begin{equation}\label{fs-def}
H_0u_j=E_ju_j\,, \qquad H_1 \tilde{u}_2=E_2 \tilde{u}_2\,,
\end{equation}
with $E_1$, $E_2$ being factorization constants.
Operator $L$ then has a nontrivial kernel space, Ker $L$, spanned by the
set of transformation functions $u_1$ and $u_2$:
\begin{equation}\label{ker}
\mathrm{Ker}\, L = \mathrm{span}\{u_1,u_2 \}\,.
\end{equation}

In the following, we will only consider self-conjugate factorization solutions,
i.e.\ solutions with a vanishing self-Wronskian W$[u,u]=0$.
The Wronskian of two matrix functions $u$, $v$ is defined as
\begin{equation}\label{Wronsk}
 \mathrm{W}[u,v](r) \equiv u^T(r)v'(r)-{u^T}'(r)v(r),
\end{equation}
leading for factorization solutions to
\begin{equation}
    \mathrm{W}[u_1,u_2](r) = u_1^T(r)\left[w_2(r)-w_1^T(r)\right] u_2(r). \label{Wronsk-spt}
\end{equation}
Hence, self-conjugate solutions correspond to symmetric superpotentials.
Solution $\tilde u_2$ then reads
\begin{equation}
\label{u2tilde}
\tilde{u}_2(r)=L_{1}u_2(r)=[w_1(r)-w_2(r)]u_2(r)
= -\left[u_1^T(r)\right]^{-1} \mathrm{W} [u_1,u_2](r)\,,
\end{equation}
where the last expression has been obtained using
\eqref{Wronsk-spt} and the symmetry of $w_1$. Using the \Sc
equation twice, one also sees that the derivative of Wronskian
\eqref{Wronsk-spt} reads
\begin{equation}\label{wr-der}
{\rm W}[u_1,u_2]'(r)=(E_1-E_2) u_1^T(r)u_2(r)\,,
\end{equation}
a relation which will be used below.

The Hamiltonians in \eqref{ir1} correspond to potentials related to each other
through superpotentials
\begin{equation}\label{pot-tr}
V_1(r)=V_0(r)-2w_1'(r)\,,\qquad V_2(r)=V_1(r)-2\tilde{w}_2'(r)\,.
\end{equation}
The sum of the two superpotentials $w_1$ and $\tilde{w}_2$
defines the two-fold superpotential $W_2$, which directly connects $V_0$ to $V_2$:
\begin{equation}\label{pt2}
W_2(r)\equiv w_1(r)+\tilde{w}_2(r)\,, \qquad V_2(r)=V_0(r)-2W'_2(r) \,.
\end{equation}
Using \eqref{w2tilde}, \eqref{Wronsk-spt} and \eqref{u2tilde}, one can rewrite $W_2$ in the compact forms
\begin{eqnarray}\label{spt2}
W_2(r) & = & (E_1-E_2)\left[w_2(r)-w_1(r)\right]^{-1} \\
& = & (E_1-E_2) u_2(r) {\mathrm W}[u_1,u_2]^{-1}(r)u_1^T(r). \label{spt2-wr}
\end{eqnarray}
As will be seen below, the second expression is more general than the first one,
as it may be used in cases where the individual superpotentials $w_1$ or $w_2$ are singular.

Similarly, expressing the second derivative of the matrix solution $\psi_0(k,r)$ from \eqref{se1}
and defining the logarithmic derivative
\begin{equation}\label{spts}
w_k(r)=\psi'_0(k,r)\psi_0^{-1}(k,r)\,,
\end{equation}
one can rewrite the action of
the second order transformation operator $L$
on $\psi_0(k,r)$,
\begin{equation}\label{fact-ch}
\psi_2(k,r)=
\left(\tilde w_2-\partial_r\right) \left(w_1-\partial_r\right) \psi_0(k,r)\,,
\end{equation}
in the following form
\begin{equation}\label{sol2-2fsusy-uns}
\psi_2(k,r)=\left[(-k^2+E_1) \mathbf{1} +W_2(r)(w_1-w_k)\right]\psi_0(k,r)\,.
\end{equation}
A more symmetric form of this formula
\begin{equation}\label{sol2-2fsusy}
\psi_2(k,r)=
\left[\left(-k^2+
\frac{E_2+E_1}{2}\right)\mathbf{1}+W_2(r)\left(\frac{w_1+w_2}{2}-w_k\right)\right]\psi_0(k,r)
\end{equation}
may also be useful.

\subsection{Main theorem}
%

Let us now particularize the above results to two consecutive SUSY trans\-formations with mutually conjugated complex matrix factorization solutions corresponding to imaginary factorization energies.
We will prove that such a second order transformation modifies the mixing parameters without affecting the eigenphase shifts.

\begin{Th}\label{T-ppt}
Consider a complex matrix solution $u$ of the coupled-channel \Sc equation \eqref{se1}-\eqref{Vzero},
with imaginary energy $E_1=k_1^2\equiv2i\chi^2$ and complex wave number $k_1=\chi(i+1)$, $\chi>0$, behaving at large distances as
\begin{equation}\label{ppt-tf-as}
u(r\rightarrow\infty) \rightarrow \left(
\begin{array}{cc} h_{l_1}\left(-k_1r\right) & \pm i h_{l_1}\left(k_1r\right)
\\ \mp i h_{l_2}\left(-k_1r\right) & h_{l_2}\left(k_1r\right) \end{array}\right),
\end{equation}
and near the origin as
\begin{equation}\label{ppt-tf-bb}
u(r\rightarrow 0)=\left(
\begin{array}{cc} a_1r^{\nu_1+1} & b_1r^{-\nu_1} \\
a_2r^{\nu_2+1} & b_2r^{-\nu_2} \end{array}\right)[1+{\rm o}(r)].
\end{equation}
The two-fold SUSY transformation defined by \eqref{ir2}-\eqref{fs-def}
with matrix facto\-rization solutions $u_1= u$, $u_2= u^*$ corresponding to the imaginary factorization constants $E_1$, $E_2=E_1^*=-2i\chi^2$ and complex wave numbers $k_1$, $k_2=\chi(i-1)$, possesses the following properties:\\
{\bf A.} The resulting potential $V_2$ defined in \eqref{pt2} is real, symmetric and regular $\forall r$.
The two-fold superpotential $W_2$ reads
\begin{eqnarray}
W_2(r) & = & 4i\chi^2\left[w^*(r)-w(r)\right]^{-1}\,,\qquad w(r)=u'(r)u^{-1}(r), \label{spt2-t} \\
    & = & 4i\chi^2u^*(r){\rm W}[u,u^*]^{-1}(r)u^T(r), \label{spt2-wr-t}
\end{eqnarray}
where only the second expression can be used when the superpotential $w$ is singular. \\
{\bf B.}
The long range behaviour of $V_2$,
\begin{equation}\label{V2infty}
V_2(r\rightarrow \infty)=\bar l(\bar l+\mathbf{1})r^{-2}+{\rm o}(r^{-2})\,,
\qquad \bar l=\diag(l_2,l_1)\,,
\end{equation}
corresponds to a re-ordering of partial waves with respect to channels.\\
{\bf C.} The scattering matrix $S_2$ of the transformed \Sc equation is expressed
from the initial scattering matrix $S_0$ as follows:
\begin{equation}\label{ppt-st}
S_2(k)=O(k)S_0(k)O^T(k)\,,
\end{equation}
where the real orthogonal matrix $O$ reads
\begin{equation}\label{ppt-uinf}
O(k) = e^{i\bar{l}\frac{\pi}{2}}
\frac{1}{\sqrt{k^4+4\chi^4}}
\left(
\begin{array}{cc} -k^2 &  \mp 2\chi^2 \\ \pm 2 \chi^2 & -k^2\end{array}\right)
e^{-il\frac{\pi}{2}}.
\end{equation}
{\bf D.} The eigenphase shifts of the transformed scattering matrix $S_2$ coincide
with the initial ones.
With the permutation
\begin{equation}\label{d2d0}
\delta_{2;1}(k)=\delta_{0;2}(k),
\end{equation}
\begin{equation}
\delta_{2;2}(k)=\delta_{0;1}(k),
\end{equation}
the mixing parameter transforms as
\begin{equation}\label{ppt-mpsd}
 \epsilon_2(k) =\epsilon_0(k) \pm (-1)^m\arctan\frac{k^2}{2\chi^2}\,.
\end{equation}
\end{Th}
\begin{proof}

First, we note that Lemma \ref{l-1} implies that solution $u$ exists.
It reads
\begin{equation} \label{ppt-tf}
 u(r)=\frac{2k_1}{i}\varphi_0(k_1,r)F_0^{-1}(k_1)\left(\begin{array}{cc} 1 & 0 \\ \mp i & 0 \end{array}\right)
+f_0(k_1,r)\left(\begin{array}{cc} 0 & \pm i \\ 0 & 1 \end{array}\right).
\end{equation}
Using (\ref{asyh}) and (\ref{ppt-tf-as}),
one may write the leading terms of the asymptotic behaviour of this factorization solution as
\begin{equation}\label{u1221as}
u(r\rightarrow\infty) \rightarrow \left(
\begin{array}{cc} e^{-ik_1 r}\left(1-\frac{i\Lambda_1}{2k_1 r}\right) &
\pm i e^{ik_1 r}\left(1+\frac{i\Lambda_1}{2k_1 r}\right) \\
\mp i e^{-ik_1 r}\left(1-\frac{i\Lambda_2}{2k_1 r}\right) &
e^{ik_1 r}\left(1+\frac{i\Lambda_2}{2k_1 r}\right) \end{array}\right).
\end{equation}
{\bf A.}
According to the choice of transformation functions and factorization
constants,
the one-fold superpotentials $w_1$ and $w_2$ are mutually complex
conjugated,
 $w_1=w$, $w_2=w^*$.
Therefore, one can use $w=u'u^{-1}$
and its complex conjugated form $w^*$
in \eqref{spts12}, \eqref{w2tilde} and \eqref{u2tilde},
thus obtaining
\begin{equation}\label{ct2c}
\tilde{w}_2(r)=\tilde{w}^*(r)=(\tilde{u}^*)'(\tilde{u}^*)^{-1}\,,
\qquad \tilde{u}^*(r)=L_{1}u^*(r)=(w-w^*)u^*\,.
\end{equation}
In this case, \eqref{spt2-t} and \eqref{spt2-wr-t} directly follow from \eqref{spt2}
and \eqref{spt2-wr}.

From \eqref{spt2-t}, it is seen that $W_2$, and thus the transformed potential
\eqref{pt2}, are real.
The symmetry of matrix $V_2$ (i.e.\ $V_2^T=V_2$) follows from the symmetry of
superpotential $w$, which can be established by considering
the self-Wronskian W$[u,u]$.
Since \eqref{wr-der} implies that this self-Wronskian is constant with respect to $r$ and \eqref{u1221as} implies that it vanishes at large distances,
${\rm W}[u,u](\infty)=0$, one has ${\rm W}[u,u](r)=0, \forall r$.
According to \eqref{Wronsk-spt}, this is equivalent to the symmetry $w^T(r)=w(r), \forall r$.

Let us now prove that $V_2$ is regular.
According to \eqref{pt2} and \eqref{spt2-wr-t},
this is the case if and only if the Wronskian W$[u,u^*]$ is invertible $\forall r$.
From \eqref{Wronsk} follows that W$[u,u^*]$ is an anti-Hermitian matrix, i.e.\
${\rm W}[u,u^*]=-{\rm W}^\dagger[u,u^*]$.
Moreover, using \eqref{wr-der}, the derivative of this Wronskian reads
\begin{equation} \label{wr-der-ct}
{\rm W}[u,u^*]'(r)=4i\chi^2u^T(r)u^*(r)\,.
\end{equation}
Its diagonal entries can thus be integrated using \eqref{ppt-tf-bb} and \eqref{u1221as} respectively.
One gets finally
\begin{equation}\label{wr-int-per-m}
\fl
{\rm W}[u,u^*](r)=\left(
\begin{array}{cc}
4i\chi^2\int\limits_0^r(|u_{11}(t)|^2+|u_{21}(t)|^2)dt &
{\rm W}_{12}[u,u^*](r)\\
-{\rm W}_{12}^* [u,u^*](r)  &
 -4i\chi^2\int\limits_r^\infty(|u_{12}(t)|^2+|u_{22}(t)|^2)dt
\end{array}\right),
\end{equation}
where $u_{ij}$ and W$_{ij}[u,u^*]$ label the entries of the factorization solution and of the Wronskian, respectively.
This result implies that ${\rm det}{\rm W}[u,u^*]>0, \forall r$,
which proves the regularity of $V_2$ stated in the theorem.
Let us stress that this proof holds even in cases where superpotential $w$ and the intermediate potential $V_1$ are singular,
which shows that expression \eqref{spt2-wr-t}, though more complictaed,
is more general than \eqref{spt2-t}.

{\bf B.}
Let us first consider the case $l_1 \neq l_2$.
From the asymptotic behaviour \eqref{u1221as},
it follows that the determinant
of the transformation solution $u$
tends to zero as $r\rightarrow\infty$ like the Laurent series
\begin{equation}\label{dtssQI}
\det u(r\rightarrow\infty) =
\frac{(\Lambda_2-\Lambda_1)}{\chi(1-i)r} + {\rm o } (r^{-2})\,.
\end{equation}
Hence, the superpotential $w$ behaves asymptotically as
\begin{equation}\label{asspt}
w(r\rightarrow\infty)=\frac{4\chi^2 r}{\Lambda_1-\Lambda_2}\left(
\begin{array}{cc} i & \pm 1\\ \pm 1& -i\end{array}\right)
+{\rm O}(1)\,,
\end{equation}
from which, using \eqref{spt2-t}, we find the asymptotic behaviour of $W_2$,
\begin{equation}\label{ppt-w2}
W_2(r\to \infty)=\frac{\Lambda_2-\Lambda_1}{2r}\left(
\begin{array}{cc} 1 & 0\\  0& -1\end{array}\right)
+{\rm o}(r^{-1})\,.
\end{equation}
It should be emphasized that from \eqref{ppt-w2} follows the
exchange of the centrifugal terms in $V_2$ with respect to $V_0$
[see \eqref{pt2}]. This effect of coupling SUSY transformations
was previously described in \cite{pupasov:09}. Note that the
scattering properties of the transformed system crucially depend
on the exchange of centrifugal terms because of the presence of
$l$-dependent factors in the $S$-matrix definition \eqref{SdefI}.

In the case of coinciding partial waves, $l_1=l_2$, \eqref{ppt-w2} is still valid but cannot be established through \eqref{asspt}:
instead, $W_2(r)$ can be calculated from the Wronskian representation \eqref{spt2-wr-t}
(see \ref{ap1}).
The fact that the two-fold superpotential vanishes at large distances faster than $r^{-1}$ implies that the centrifugal tails are not affected by the SUSY transformations.


{\bf C.}
To establish the modification of the scattering matrix,
we have to look at the way the Jost solutions and
the regular solutions transform in the two-fold transformation.

Once again, let us start with the simpler case $l_1\neq l_2$.
Without loss of generality we may apply
the general transformation of solutions \eqref{sol2-2fsusy}
to the Jost solution,
which now takes the form
\begin{equation}\label{sol2-2fsusy-t}
L f_0(k,r)=
\left[-k^2 \mathbf{1} +W_2(r)\left(\frac{w+w^*}{2}-w_k\right)\right]f_0(k,r)
\equiv U(k,r)f_0(k,r).
\end{equation}

As we will see below, the matrix $U_{\infty}(k)=\lim_{r\to\infty}U(k,r)$
determines the transformed Jost and scattering matrices.
Using \eqref{asspt}, \eqref{ppt-w2} and the fact that
$W_2 w_k$ vanishes at large distances,
one obtains a simple expression for this matrix,
\begin{equation}\label{uinf}
U_{\infty}(k)=\left(
\begin{array}{cc} -k^2 & \mp 2 \chi^2 \\ \pm 2 \chi^2 & -k^2\end{array}\right).
\end{equation}
From the dominant term of \eqref{jsas} and \eqref{asyh},
it follows that the function
\begin{equation}\label{josttr}
f_2(k,r)=Lf_0(k,r)U_{\infty}^{-1}(k)
\end{equation}
is the transformed Jost solution.

As in the previous part, the case $l_1=l_2$ requires additional attention
since the product $W_2(w+w^*)$ gives at large distances the uncertainty $0\cdot\infty$. Again we use the
Wronskian representation \eqref{spt2-wr-t} of the two-fold superpotential $W_2$ and
the asymmetrical form of transformation \eqref{sol2-2fsusy-uns} thus obtaining
\begin{equation}
L f_0(k,r)
\mathop{\rightarrow}_{r\rightarrow\infty}
\left[(-k^2+2i\chi^2)\mathbf{1}
+4i\chi^2u^*{\rm W}[u,u^*]^{-1}{u^T}'\right]f_0(k,r)\,.
\end{equation}
Using \eqref{app-uas} and \eqref{app-Wm1} in this expression leads to the same matrix $U_{\infty}(k)$ as in \eqref{uinf}.

Let us now find how the SUSY transformation modifies
the behaviour of the potential at the origin.
From \eqref{ppt-tf-bb},
one gets
\begin{equation}
 \det u(r\rightarrow 0) \rightarrow a_1 b_2 r^{\nu_1-\nu_2+1} - a_2 b_1 r^{\nu_2-\nu_1+1},
\end{equation}
which suggests that the discussion will depend on the relative values of $\nu_1$ and $\nu_2$.

For $\nu_2=\nu_1$, excluding the case $a_1 b_2=a_2 b_1$ (which requires higher order expansions), one can expand the superpotential $w(r)$
in a Laurent series near $r=0$,
\begin{equation} \label{spt0}
\fl
 w(r\rightarrow 0) = \frac{1}{(a_1b_2-a_2b_1)r}
\left( \begin{array}{cc}
    a_1b_2 (\nu_1+1) +a_2b_1 \nu_1 & -a_1 b_1 (2 \nu_1+1) \\
    a_2 b_2 (2 \nu_2+1) & -a_2b_1(\nu_2+1)-a_1 b_2 \nu_2
       \end{array} \right) +\mathrm{o}(1),
\end{equation}
which implies with \eqref{spt2-t} that the lowest-order term in $W_2$ is linear in $r$.
Consequently, \eqref{pt2} implies that the singularity indices are not
modified by the two-fold SUSY transformation.
Note however that \eqref{pot-tr} implies that the intermediate potential $V_1$
displays in general off-diagonal singular terms at the origin.

For $\nu_2>\nu_1$, one gets instead of \eqref{spt0}
\begin{equation}\label{spt1-bb}
w(r\rightarrow 0)=\frac{1}{r}\left(
\begin{array}{cc}
\nu_1+1 & 0 \\
0 & -\nu_2 \end{array}\right) + {\rm o}(1).
\end{equation}
To find the behaviour of $W_2$ at the origin,
a higher-order expansion would thus be necessary.
It is simpler in this case to study the two first-order transformations separately.
From \eqref{pot-tr} and \eqref{spt1-bb}, we conclude that
the intermediate potential $V_1$ has the following  singularity indices
$\nu\rightarrow\tilde\nu=\diag(\nu_1+1,\nu_2-1)$.
For $\nu_2<\nu_1$, one gets $\nu\rightarrow\tilde\nu=\diag(\nu_1-1,\nu_2+1)$ by symmetry.

Let us now analyze the behaviour of
the transformation function $\tilde u^*=L_1u^*$ which
determines operator $L_2$.
Using \eqref{ct} and \eqref{spt1-bb} [or \eqref{spt0} when $\nu_1=\nu_2$] one can find that a regular/singular
vector solution transforms into a regular/singular vector
solution of the new equation. Such transformations are called conservative SUSY transformations \cite{sparenberg:06}.
As a result the behaviour of $\tilde u^*$ near the origin
is given by the conjugate of \eqref{ppt-tf-bb}
with different values of constants
$a_{1,2}^*$ and $b_{1,2}^*$, i.e.,
$a_{1,2}^*\rightarrow \tilde a_{1,2}^*$
and $b_{1,2}^* \rightarrow \tilde b_{1,2}^*$,
and shifted singularity indices $\tilde \nu=\diag(\nu_1+1,\nu_2-1)$
(to fix ideas, we consider the case $\nu_2>\nu_1$)
\begin{equation}\label{ppt-tf-bb-int}
\tilde u^*(r\rightarrow 0)=\left(
\begin{array}{cc} \tilde a_1^*r^{\nu_1+2} & \tilde b_1^*r^{-\nu_1-1} \\
\tilde a_2^*r^{\nu_2} & \tilde b_2^*r^{-\nu_2+1} \end{array}\right) [1+\mathrm{o}(r)].
\end{equation}

We have to split the discussion into two subcases, once again.
For $\tilde\nu_2=\tilde\nu_1$, i.e. $\nu_2=\nu_1+2$, an equation similar to \eqref{spt0} implies that $\tilde w^*$ behaves
like $r^{-1}$ multiplied by a non-diagonal matrix close to the origin.
Consequently, the final potential $V_2$ will be unphysical in general, with non-diagonal singular terms at the origin;
therefore, we will not consider this case any further.
For $\tilde\nu_2>\tilde\nu_1$, i.e.\ $\nu_2>\nu_1+2$, the same reasoning as above implies that the transformed potential $V_2$ has the
following singularity indices:
$\tilde\nu\rightarrow \bar\nu=\diag(\tilde\nu_1+1, \tilde\nu_2-1)=
\diag(\nu_1+2,\nu_2-2)$.
Finally, for $\tilde\nu_2<\tilde\nu_1$,
which  is the case for $\nu_2=\nu_1+1$,
the second transformation restores the
initial singularity indices
$\tilde\nu\rightarrow  \bar\nu=\diag(\tilde\nu_1-1,\tilde\nu_2+1)=\diag(\nu_1,\nu_2)$.

The modification rules for the singularity indices of the potential
may thus be summarized as follows in the physical cases:
\begin{equation}
(\nu_1,\nu_1)\stackrel{L}{\longrightarrow}(\nu_1,\nu_1)\,,
\end{equation}
\begin{equation}
(\nu_1,\nu_1+1)\stackrel{L}{\longrightarrow}(\nu_1,\nu_1+1)\,,
\end{equation}
\begin{equation}
(\nu_1,\nu_1+m)\stackrel{L}{\longrightarrow}(\nu_1+2,\nu_1+m-2)\,,\qquad m>2.
\end{equation}
From here it is seen that in all cases $\Tr \nu=\Tr \bar\nu$.

We are now ready to construct the regular solution of the transformed \Sc equation.
For $\nu_2\neq\nu_1$ superpotentials
$w$ and $\tilde w^*$ have the structure given by \eqref{spt0} or \eqref{spt1-bb}
depending on the singularity indices. Therefore the first-order transformations $L_1$ and $L_2$
are conservative.
Thus, the result of
the two-fold SUSY transformation applied to $\varphi_0(k,r)$ in the most general form
can be written  as follows
\begin{equation}\label{regst-n}
L\varphi_0(k,r)=\varphi_2(k,r)U_0(k)\,,
\end{equation}
where $U_0$ is a constant matrix with respect to $r$.
Matrix $U_0(k)$ is invertible $\forall k\neq k_{1,2}$,
which can be seen from \eqref{ker}.
In the case $\nu_2=\nu_1$, the conservativeness of the two-fold
SUSY transformation can be established by considering
\eqref{sol2-2fsusy-t} where $\psi_0$ is replaced by a regular solution.
Note that $\varphi_{0,2}(k,r)=\varphi_{0,2}(-k,r)$;
therefore, matrix $U_0$ is an even matrix function of wave number $k$,
$U_0(k)=U_0(-k)$.
The precise value of $U_0$ is not important for the following.

Applying operator $L$ to the relation \eqref{rajs} between the Jost solutions and the regular solution,
one obtains with \eqref{josttr} and \eqref{regst-n}
\begin{equation}\label{fckr1-2fsusy}
\varphi_2(k,r)U_0(k)=
\frac{i}{2k}\left[f_2(-k,r)U_{\infty}(-k)F_0(k)-f_2(k,r)U_{\infty}(k)F_0(-k)\right].
\end{equation}
The transformed Jost matrix thus reads
\begin{equation}\label{tjm2c-2fsusy}
F_2(k)=U_{\infty}(-k)F_0(k)U_0^{-1}(k)\,.
\end{equation}
The transformation of the scattering matrix then follows from its definition \eqref{SdefI},
\begin{equation}\label{ppt-st-pr}
S_2(k)=e^{i\bar{l}\frac{\pi}{2}}
U_{\infty}(k)e^{-il\frac{\pi}{2}}S_0(k)e^{-il\frac{\pi}{2}}%
U_{\infty}^{-1}(k)e^{i\bar{l}\frac{\pi}{2}}\,,
\end{equation}
and is equivalent to \eqref{ppt-st} and \eqref{ppt-uinf}.
Note that the transformed $S$-matrix does not depend on $U_0$.
To prove that matrix $O$ is real and orthogonal,
one has to remember that $l_1$, $l_2$, $\bar l_1$, $\bar l_2$ all have the same parity,
as implied by \eqref{l2l1} and \eqref{V2infty}.
If written like in definition \eqref{par-def},
matrix $O$ corresponds to a rotation angle
\begin{equation}
 \mp(-1)^m\arctan\frac{2 \chi^2}{k^2}=\frac{\pi}{2}\pm(-1)^m \arctan\frac{k^2}{2\chi^2}.
\label{epsadd}
\end{equation}

{\bf D.}
Diagonalizing $S_2$ in the same way as $S_0$ in \eqref{S0diag},
\begin{equation}
R_2^T(k) S_2(k) R_2(k)= \diag \left(e^{2i\delta_{2;1}(k)},e^{2i\delta_{2;2}(k)}\right),
\end{equation}
and taking into account that matrices $R_0$, $O$ and $R_2=O R_0$ all belong to $SO(2)$,
one sees that $S_0$ and $S_2$ have the same eigenvalues.
The mixing angle of $S_2$ is given by the sum of $\epsilon_0$ and \eqref{epsadd}.
Inverting the order of these eigenvalues (see discussion following \eqref{par-def}),
one gets \eqref{d2d0} and \eqref{ppt-mpsd},
i.e.\ a modification of mixing parameter vanishing at zero energy, $\epsilon_2(0)-\epsilon_0(0)=0$.
\end{proof}




\subsection{Iteration}

Let us finally note that the transformed potential $V_2$
can be used as a starting point for a next eigenphase preserving transformation.
This means that the two-fold SUSY
transformation considered above can be iterated as long as
desirable.
A chain of $n$ such transformations over the initial
potential $V_0$ will lead to the following mixing parameter:
\begin{equation}\label{ppt-mpsd-n}
 \epsilon_{2n}(k) =\epsilon_0(k) \pm (-1)^m \sum\limits_{j=0}^n
 \arctan\frac{k^2}{2\chi_j^2}
\end{equation}
leaving the eigenphase shifts unchanged.


\section{Example}
Let us consider a simple example where the eigenphase preserving
SUSY transformation is applied to an $s-d$ diagonal potential with
the following scattering matrix:
\begin{equation}\label{ism}
S_0(k)=\diag\left(1,\frac{(k+i \k_1)(k+i \k_2)}{(k-i \k_1)(k-i \k_2)}\right).
\end{equation}
The first channel corresponds to the $d$ wave and the second channel corresponds to the $s$ wave.
The corresponding potential reads
\begin{equation}\label{Vd1}
V_0(r)=\diag\left(\frac{6}{r^2},
-2 \bigg(\ln {\mathrm W}\left[v_1,v_2\right](r)\bigg)''\right).
\end{equation}
The $d$-wave potential is purely centrifugal,
while the $s$-wave potential is obtained from the zero potential by a second order one-channel SUSY transformation
 with the factorization solutions
$v_1(r)=\sinh(\k_1r)$ and $v_2(r)=\sinh(\k_2r)$.
This $s$-wave potential has no bound state but a singular repulsive core at the origin \cite{sparenberg:97a}.
Potential $V_0$ is thus characterized
by the singularity and centrifugal indices
\begin{equation}
\nu=\diag(2,2)\,,\quad l=\diag(2,0)\,.
\end{equation}
The Jost solution corresponding to potential $V_0$ reads
\begin{equation}\label{js-0}
f_{0}(k,r)=\diag\bigg(f_{0d}(k,r),f_{0s}(k,r)\bigg),
\end{equation}
where
\begin{eqnarray}\label{js02}
f_{0d}(k,r) & = & {\rm e}^{ikr}\left(1+\frac{3i}{kr}-\frac{3}{(kr)^2}\right)\,, \\
f_{0s}(k,r) & = & \left(\frac{\tilde v'_2(r)}{\tilde v_2(r)}-\p_r\right)
\left(\frac{v_1'(r)}{v_1(r)}-\p_r\right){\rm e}^{ikr}N_{1}N_{2}\,,
\end{eqnarray}
with $\tilde v_2=[(\ln v_1)'-\p_r]v_2$ and the normalization constants $N_{j}=(ik-\kappa_j)^{-1}$.
The regular solution $\varphi_0$ is expressed
from \eqref{rajs} with the Jost matrix
\begin{equation}\label{pejf1}
F_0(k)=\diag\left(1,-N_1 N_2\right).
\end{equation}

Using these expressions for the Jost and regular solutions,
one may construct with \eqref{ppt-tf} a transformation solution $u$ with asymptotics \eqref{ppt-tf-as} and
\eqref{ppt-tf-bb}, according to Lemma \ref{l-1}. The eigenphase
preserving transformation described in Theorem \ref{T-ppt} leads
to a singular potential $V_2$ without bound state and with
\begin{equation}
\bar{\nu}=\diag(2,2)\,,\quad \bar{l}=\diag(0,2)\,.
\end{equation}
The eigenphase shifts of the transformed $S$-matrix coincide with
the initial eigenphase shifts,
\begin{eqnarray}\label{tcps}
\d_{s}(k) & = & -\arctan\frac{k}{\k_1}-\arctan\frac{k}{\k_2}\,,\\
\d_{d}(k) & = & 0\,.
\end{eqnarray}
The mixing angle is given by \eqref{ppt-mpsd} with $\epsilon_0=0$.
In this case, different signs in \eqref{ppt-mpsd} correspond to
different signs in the coupling interaction $V_{2sd}\leftrightharpoons -V_{2sd}$.

\begin{figure}
\begin{center}
\epsfig{file=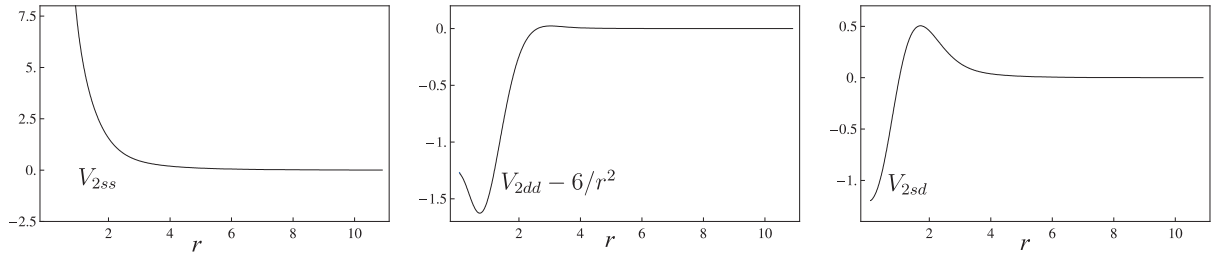, width=16cm} \caption{
Entries of the exactly solvable potential matrix $V_2$ obtained from the uncoupled potential \eqref{Vd1} with parameters $\k_1=0.232\,,\k_2=0.944$ by application of the
eigenphase preserving transformation with $\chi=1.22$.
\label{figSStrpt}}
\end{center}
\end{figure}

The transformed potential $V_2$ with the following parameters:
\begin{equation}
\k_1=0.232\,,\qquad \k_2=0.944\,,\qquad \chi=1.22\,,
\end{equation}
is shown in Figure \ref{figSStrpt}
[for definiteness we have chosen the positive sign in \eqref{ppt-mpsd}].
The main reason to consider this example is that
it illustrates the same scattering matrix as the one
obtained by Newton and Fulton
in \cite{newton:57}.
The Newton-Fulton potential
differs from the potential constructed above
because it has one bound state.
This difference can in principle be
eliminated by the well known technique of the coupled-channel phase-equivalent
bound state addition \cite{Leeb}.

\section{Conclusion}

In this paper, we have introduced an ``eigenphase preserving'' two-fold SUSY
trans\-for\-mation for the two-channel \Sc equation, i.e.\ a transformation that alters the mixing parameter between channels without modifying the eigenphase shifts.
Chains of such transformations lead to coupling between
channels in the scattering matrix which correspond to nontrivial
$k$-dependences of the mixing angle \eqref{ppt-mpsd-n}.
With a reasonably small number of parameters,
such mixing angles are probably able to fit experimental data,
in a similar way to the usual phase shift fitting used in one-channel SUSY inversion \cite{sparenberg:97,samsonov:03}.
Combining both techniques,
we obtain a complete method of coupled-channel scattering data inversion based on SUSY transformations.
As a first application of this method,
we plan to invert the two-channel neutron-proton scattering data,
hence improving the result of \cite{newton:57}.

We also plan to study the following questions,
raised by the present work.
How do bound states transform under this eigenphase preserving transformation?
How to construct a similar transformation for an arbitrary number of coupled
channels?
Do other forms of eigenphase preserving transformations exist?
How will the presence of the Coulomb interaction modify the properties of the
eigenphase preserving SUSY transformation?

\ack
AP and BFS are supported by the Russian Federal Agency of Education
under contract No.\ P1337, the President of Russia under the grant No.\ SS-871.2008.2
and the Russian Science and Innovations Federal Agency under contract No.\ 02.740.11.0238.
AP is supported by the Russian "Dynasty" foundation and the Russian Federal Agency of Education under contract No.\ P2569.
This text presents research results of the Belgian Research Initiative on eXotic nuclei (BriX), program P6/23 on interuniversity attraction poles of the Belgian Federal Science Policy Office.

\appendix
\section{\label{ap1}}
Let us calculate asymptotics \eqref{ppt-w2} using the Wronskian representation \eqref{spt2-wr-t}.
This allows us to avoid manipulations with singular quantities which
appear in \eqref{asspt} when $l_1=l_2$. It is convenient to rewrite the asymptotic
behaviour of the transformation solution in the form
\begin{equation}
\fl
u(r\rightarrow \infty)\to \left(2 Q_{\mp}-\frac{i}{\xi_1}\Lambda Q_{\mp}\sigma_z \right){\rm e}^{-i\xi_1\sigma_z}\,,
\qquad Q_{\mp}=(\mathbf{1}\mp\sigma_y)/2\,,
\qquad \xi_1=k_1r,
\label{app-uas}
\end{equation}
where $\Lambda=\diag(\Lambda_1,\Lambda_2)$,
$\sigma_x$, $\sigma_y$ and $\sigma_z$ are the Pauli matrices,
and the projection matrices $Q_{\mp}$ satisfy
\begin{equation}\label{qtqq}
Q_{\pm}^T=Q_{\mp}\,,
\qquad Q_{\pm}Q_{\mp}=0\,,\qquad Q_{\pm}^2=Q_{\pm}\,,
\end{equation}
\begin{equation}\label{qssq}
Q_{\pm}\sigma_z=\sigma_zQ_{\mp}\,,
\qquad Q_{\pm}\sigma_x=\sigma_xQ_{\mp}.
\end{equation}
Here and in what follows we will only retain terms of order $r^{-1}$ or lower.
Let us first calculate
the Wronskian asymptotics at large distances.
Definition \eqref{Wronsk} leads to
\begin{equation}
{\rm W}(r\to\infty)\rightarrow 4i\chi {\rm e}^{-i\xi_1\sigma_z}(\sigma_z\pm\sigma_x)\left[
1-\frac{(1-i)}{4\chi r}(\Lambda_1+\Lambda_2)\sigma_zQ_{\mp}
\right]{\rm e}^{i\xi_1^*\sigma_z}\,,
\end{equation}
which can be inverted (up to $r^{-1}$) to give
\begin{eqnarray}
{\rm W}^{-1} (r\to\infty)& \rightarrow & \frac{1}{8i\chi} {\rm e}^{-i\xi^*_1\sigma_z}\left[
1+\frac{(1-i)}{4\chi r}(\Lambda_1+\Lambda_2)\sigma_zQ_{\mp}
\right](\sigma_z\pm\sigma_x){\rm e}^{i\xi_1\sigma_z} \\
& =& \frac{1}{8i\chi} {\rm e}^{-i\xi_1^*\sigma_z}\left[
\sigma_z\pm\sigma_x
+\frac{1}{2\chi r}(\Lambda_1+\Lambda_2)Q_{\pm}
\right]{\rm e}^{i\xi_1\sigma_z}\,.
\label{app-Wm1}
\end{eqnarray}
We can now calculate the two-fold superpotential up to $r^{-1}$
\begin{eqnarray}
\fl W_2 & = & 4i\chi^2 u^*{\rm W}^{-1}u^T\\
\fl &  \to  & \chi \left(\frac{i}{\xi_1^*}\Lambda Q_{\pm}\sigma_z(\sigma_z\pm\sigma_x)Q_{\pm}
+\frac{1}{\chi r}(\Lambda_1+\Lambda_2)Q_{\pm}-
\frac{i}{\xi_1} Q_{\pm}(\sigma_z\pm\sigma_x)\sigma_zQ_{\pm}\Lambda
\right),
\end{eqnarray}
where \eqref{qtqq} and \eqref{qssq} have been used.
To further simplify this expression, we also use
the decomposition ${\Lambda={\bf 1}(\Lambda_1+\Lambda_2)/2+\sigma_z(\Lambda_1-\Lambda_2)}$,
which leads finally to
\begin{equation}
W_2(r\to\infty)  \to  \frac{1}{2r}(\Lambda_2-\Lambda_1)\sigma_z\,.
\end{equation}
This expression is valid for any $l_1$ and $l_2$;
it is thus also valid for the case of coinciding partial waves.

\end{document}